\documentclass[11pt,a4paper]{article}
\usepackage[utf8]{inputenc}
\usepackage{amsmath}
\usepackage{amsthm}
\numberwithin{equation}{section}
\usepackage{amsfonts}
\usepackage{mathtools}
\usepackage{mathrsfs}
\usepackage{mathpazo}
\usepackage{multirow}
\usepackage{amssymb}
\usepackage{graphicx}
\usepackage{ifpdf}
\usepackage{braket}
\usepackage[dvipsnames]{xcolor}

\usepackage{cite}
\usepackage[bookmarks=true,colorlinks=true,linkcolor=black,citecolor=orange,urlcolor=orange,bookmarksnumbered]{hyperref}

\usepackage[left=2.50cm, right=2cm, top=2cm, bottom=3cm]{geometry}
\begin{document}
\thispagestyle{empty}

\begin{center}

\vspace{2.0truecm}

{\Large \bf A heterotic integrable deformation of the principal chiral model}

\vspace{2.0truecm}

{David Osten}

\vspace{0.5truecm}

{\em Institute for Theoretical Physics (IFT), \\
University of Wroc\l aw \\
pl. Maxa Borna 9, 50-204 Wroc\l aw, Poland
}

\vspace{0.5truecm}

{{\tt david.osten@uwr.edu.pl}}

\vspace{0.5truecm}
\end{center}

\begin{abstract}
A novel classically integrable model is proposed. It is a deformation of the two-dimensional principal chiral model, embedded into a heterotic $\sigma$-model, by a particular heterotic gauge field. This is inspired by the bosonic part of the heterotic $\sigma$-model and its recent Hamiltonian formulation in terms of O$(d,d+n)$-generalised geometry in \cite{Hatsuda:2022zpi}. Classical integrability is shown by construction of a Lax pair and a classical $\mathcal{R}$-matrix. Latter is almost of the canonical form with twist function and solves the classical Yang-Baxter equation.
\end{abstract}

\tableofcontents

\section{Introduction}
One of the main impacts of the study of integrable model in high energy physics is the connection of quantities of integrable Green-Schwarz superstring $\sigma$-models in supercoset Anti de Sitter backgrounds and those of dual integrable conformal field theories \cite{Beisert:2010jr,Bombardelli:2016rwb} in context of the AdS/CFT-duality. In the last years, some effort has been spent to generalise this to other, less symmetric settings. On the world-sheet side of the duality, new integrable models were found that are deformations of such (semi)symmetric (super)cosets, namely Yang-Baxter deformations \cite{Klimcik:2002zj,Klimcik:2008eq,Delduc:2013qra}, $\lambda$-\cite{Sfetsos:2013wia} and WZ-term \cite{Klimcik:2017ken} deformations and their interplay \cite{Delduc:2014uaa}, or generalisations to non-symmetric cosets \cite{Young:2005jv,Hoare:2021dix,Osten:2021opf}. So far, this survey was focussing on conventional two-dimensional non-linear $\sigma$-models, i.e. the embedding of a two-dimensional world-sheet in some $d$-dimensional target space. See \cite{Hoare:2021dix,Driezen:2021cpd} for two recent reviews.

This article aims to slightly step out of this fairly well explored context and suggest that also the heterotic string \cite{Green:1984sg,Gross:1985fr,Sen:1985qt,Sen:1985eb} might have interesting integrable models. In context of this paper, the bosonic part of a heterotic string $\sigma$-model is a $d$-dimensional non-linear $\sigma$-model, characterised by a metric and a $B$-field, together with $n$ chiral currents, characterised by a gauge field $A$ and a given $n$-dimensional gauge group.

\paragraph{A heterotic $\mathcal{E}$-model.} Recently, progress in understanding the space of two-dimensional classically integrable $\sigma$-models was made using two quite different constructions: in terms of as affine Gaudin models \cite{Vicedo:2017cge,Lacroix:2018njs}, coming from a four-dimensional Chern-Simons theory \cite{Costello:2017dso,Costello:2018gyb,Bittleston:2020hfv,Lacroix:2020flf,Levine:2023wvt,Cole:2023umd}, or as $\mathcal{E}$-models \cite{Klimcik:1995ux,Sfetsos:1999cc,Demulder:2018lmj,Osten:2019ayq,Hassler:2020xyj,Borsato:2021gma,Borsato:2021vfy}. Latter does not directly clarify the origin of the integrable structure but its data relates directly to the $\sigma$-model target space geometry in the generalised flux formulation. Concretely, an $\mathcal{E}$-model is defined in terms of O$(d,d)$-covariant world-sheet currents $J_A$ with Poisson brackets -- the so-called current algebra --
\begin{equation}
	\{ J_A(\sigma) , J_B(\sigma^\prime) \} = \eta_{AB} \delta^\prime(\sigma-\sigma^\prime) - {F^C}_{AB}(\sigma) J_C(\sigma) \delta(\sigma-\sigma^\prime), \label{eq:CurrentAlgebraOdd}
\end{equation}
where $A = 1,...,2d$, $d$ is the dimension of the target space, $F_{ABC}$ are the so-called generalised fluxes, which are constant for an $\mathcal{E}$-model, and $\eta_{AB}$ is the O$(d,d)$-invariant metric, together with a Hamiltonian $H = \frac{1}{2} \int \mathrm{d}\sigma \ \delta^{AB} J_A J_B$.

Similar to that, O$(d,d)$ generalised geometry appeared first in the Hamiltonian formulation and the analysis of the current algebra of the world-sheet string \cite{Tseytlin:1990va,Tseytlin:1990nb,Siegel:1993th,Siegel:1993bj} before it was considered as a tool for supergravity \cite{Hull:2009mi,Zwiebach:2011rg,Aldazabal:2013sca,Hohm:2013bwa,Geissbuhler:2013uka,Plauschinn:2018wbo,Lescano:2021lup}. 

For the heterotic string the duality group is O$(d,d+n)$ \cite{Maharana:1992my,Schwarz:1993mg}. Heterotic supergravity was phrased in terms of $O(d,d+n)$ generalised geometry \cite{Hohm:2011ex,Grana:2012rr,Bedoya:2014pma,Lescano:2021lup}. World-sheet formulations of the heterotic string related to O$(d,d+n)$ generalised geometry have been suggested in \cite{Siegel:1993xq}, and recently \cite{Hatsuda:2022zpi}. This is the latest addition to the program of duality covariant formulations of world-volume theories either via actions \cite{Sakatani:2016sko,Blair:2017hhy,Sakatani:2017vbd,Arvanitakis:2018hfn,Blair:2019tww,Sakatani:2020umt} or via Hamiltonian formulations \cite{Duff:1990hn,Berman:2010is,Hatsuda:2012vm,Duff:2015jka,Sakatani:2020iad,Sakatani:2020wah,Osten:2021fil,Osten:2023iwc}. Following \cite{Hatsuda:2022zpi}, a heterotic $\mathcal{E}$-model is defined in section \ref{chap:Review}. 

\paragraph{A new integrable deformation of the principal chiral model.} The new integrable model proposed here is a deformation of the principal chiral model, embedded into a heterotic string $\sigma$-model. The deformation is characterised by a heterotic gauge field:
\begin{equation}
A^\alpha_a = \epsilon \mathcal{A}^\alpha_a, \qquad \text{with} \quad \mathcal{A}^\alpha_a \mathcal{A}^\beta_b \kappa_{\alpha \beta} = \kappa_{ab},
\end{equation}
which is constant in flat indices $a,b,c$. Here $\epsilon$ is the deformation parameter and $\kappa_{ab}$, $\kappa_{\alpha\beta}$ are the Killing forms on $\mathfrak{g}$ (the Lie algebra from the principal model) and a symmetric non-degenerate bilinear form on $\mathfrak{u}{(1)}^n$ (the abelian gauge algebra) respectively. The Lax pair and the classical $\mathcal{R}$-matrix, to show Hamiltonian integrability in Maillet's formalism \cite{Maillet:1985ec,Maillet:1986}, are computed in section \ref{chap:Adeformation}. It also shown that this deformation can be combined with the introduction of a WZ-term with arbitrary coefficient while maintaining classical integrability.

\section{The classical bosonic heterotic string in the Hamiltonian formalism} \label{chap:Review}
In a slight stretch of the concept, a (classical) bosonic heterotic string here will be a $d$-dimensional non-linear $\sigma$-model together with $n$ (abelian) chiral currents. The $d$ coordinate fields $x$, their canonical duals $p$ and $n$ chiral currents $k$ can be arranged into an O$(d,d+n)$-vector:
\begin{equation}
	\mathcal{P}_\mathcal{M} = \left( p_m(\sigma) , k^\mu(\sigma) , \partial x^m (\sigma) \right).
\end{equation}
Latin indices $k,l,m,... = 1 , ... ,d$ will be reserved for the $d$-dimensional coordinates, Greek letters $\kappa,\lambda,\mu, ... = 1,...,n$ for the $n$-dimensional coordinates. Latter can be raised and lowered using $\kappa_{\mu \nu}$, the Killing metric on $\mathfrak{u}(1)^n$. The canonical Poisson brackets of these fields can (up to topological terms) also be written in an O$(d,d+n)$-covariant way:
\begin{equation}
	\{ \mathcal{P}_\mathcal{M}(\sigma) , \mathcal{P}_\mathcal{N}(\sigma^\prime) \} = \eta_{\mathcal{M} \mathcal{N}} \delta^\prime (\sigma - \sigma^\prime)
\end{equation}
with the O$(d,d+n)$-invariant metric
\begin{equation}
	\eta_{\mathcal{M} \mathcal{N}} = \left( \begin{array}{ccc} 0 & 0 & \delta_m^n \\ 0 & \kappa_{\mu \nu} & 0 \\ \delta^m_n & 0 & 0
	\end{array} \right)
\end{equation}
which is also used to raise and lower the $\mathcal{K},\mathcal{L},\mathcal{N},...$-indices. The action on functions of these currents is $\{ \mathcal{P}_\mathcal{M} , f(x) \} = - \partial_\mathcal{M} f(x) = ( - \partial_m f(x) , 0 ,0 )$. The derivative $\partial_\mathcal{M}$ is subject to the section condition $\eta^{\mathcal{M} \mathcal{N}} \partial_\mathcal{M} \ast \partial_\mathcal{N} \ast = 0 = \eta^{\mathcal{M} \mathcal{N}} \partial_\mathcal{M} \partial_\mathcal{N} \ast$, which is an O$(d,d+n)$-covariant way to express that the functions only depend on $x$.

The coupling to the bosonic part of heterotic supergravity, namely a $d$-dimensional metric $G$ and two-form gauge field $B$ and the heterotic gauge field $A$ can be encoded in a Hamiltonian that is quadratic in the currents $\mathcal{P}_\mathcal{M}$ and contains the couplings in the so called generalised metric $\mathcal{H}_{\mathcal{M} \mathcal{N}}$:
\begin{align}
	H = \frac{1}{2} \int \mathrm{d} \sigma \ \mathcal{H}^{\mathcal{M}\mathcal{N}} (G,B,A) \mathcal{P}_{\mathcal{M}}(\sigma) \mathcal{P}_{\mathcal{N}} (\sigma). \label{eq:HamiltonianMetricFrame}
\end{align}
The spatial diffeomorphism constraint also takes a simple form, $0 = \mathcal{P}_{\mathcal{M}} \mathcal{P}_{\mathcal{N}} \eta^{\mathcal{M} \mathcal{N}}$, but will not play a role in the further discussion. The explicit generalised metric $\mathcal{H}(G,B,A)$ and the connection to a Lagrangian picture, and also the generalisation to non-abelian currents has been discussed in detail in \cite{Hatsuda:2022zpi}.

For the purpose of this paper the generalised flux formulation is better suited. To this end, we diagonalise the generalised metric $\mathcal{H}_{\mathcal{A} \mathcal{B}} = {E_\mathcal{A}}^\mathcal{M} {E_\mathcal{B}}^\mathcal{N} \mathcal{H}_{\mathcal{M} \mathcal{N}}$ with a generalised vielbein ${E_\mathcal{A}}^\mathcal{M}$ and some suitable constant $\mathcal{H}_{\mathcal{A} \mathcal{B}}$. The coupling to the background is encoded in the Poisson brackets of the currents $J_\mathcal{A} = {E_\mathcal{A}}^\mathcal{M} \mathcal{P}_\mathcal{M}$:
\begin{equation}
	\{ J_\mathcal{A}(\sigma) , J_\mathcal{B}(\sigma^\prime) \} = \eta_{\mathcal{AB}} \delta^\prime(\sigma-\sigma^\prime) - {F^\mathcal{C}}_{\mathcal{A}\mathcal{B}}(\sigma) J_\mathcal{C}(\sigma) \delta(\sigma-\sigma^\prime) \label{eq:CurrentAlgebraGeneral}
\end{equation}
via the so-called generalised fluxes $F_{\mathcal{A}\mathcal{B}\mathcal{C}} = 3 \partial_\mathcal{[A} {E_\mathcal{B}}^\mathcal{N} {E_\mathcal{C]N}}$. In the particular case that these $F_{\mathcal{ABC}}$ can be chosen to be constant, we would call this construction \textit{heterotic $\mathcal{E}$-model}, in analogy to the construction for ordinary $2d$ $\sigma$-models and O$(d,d)$-generalised geometry. For a generic heterotic supergravity background these fluxes are not constant. 

In the standard (geometric) parametrisation of the background in terms of metric $G$, two-form gauge field $B$ and heterotic gauge field $A$ the generalised vielbein is \cite{Hohm:2011ex}
\begin{equation}
	{E_\mathcal{A}}^\mathcal{M} = \left( \begin{array}{ccc} \delta_a^b & A_a^\alpha & B_{ab} - \frac{1}{2} A_{\alpha a} A^\alpha_b \\ 0 & \delta_\alpha^\beta & - A_{\alpha b} \\
	0 & 0 & \delta^a_b
	\end{array} \right) \left( \begin{array}{ccc} {e_b}^m & 0 & 0 \\ 0 & {\delta_\beta}^\mu & 0 \\ 0 & 0 & {e^b}_m	
	\end{array} \right) \label{eq:GeneralisedVielbeinGeometric}
\end{equation}
with ${e_a}^m {e_b}^n G_{mn} = \kappa_{ab}$, which will be used to raise and lower $a,b,...$ indices. Then, the non-vanishing components of $F_{\mathcal{ABC}}$ are:
\begin{align}
H_{abc} &= F_{abc} = \left[ \partial_a B_{bc} + {f^d}_{ab} \left( B_{cd} - \frac{1}{2} A^\alpha_{c} A_{d \alpha} \right) + (\partial_{[a} A_{b]}^\alpha) A_{\alpha c} \right] + \text{cyclic permutations of }abc \nonumber \\
{f^c}_{ab} &= {F^c}_{ab} = 2 {e_n}^c \partial_{[a} {e_{b]}}^n  \label{eq:GeneralisedFluxes} \\
{F^\gamma}_{ab} &= 2 \partial_{[a} A_{b]}^\gamma - {f^c}_{ab} A_c^\gamma. \nonumber
\end{align}
In hindsight of what is to come, we write $J_\mathcal{A} = (J_a , J^\alpha, J^a) = (j_{0a} , k^\alpha , j_1^a)$. The subscripts $0,1$ refer to the world-sheet coordinates $\tau, \sigma$. In this decomposition the current algebra \eqref{eq:CurrentAlgebraGeneral} is
\begin{align}
\{ j_{0a}(\sigma) , j_{0b}( \sigma^\prime ) \} &= \left( -{f^c}_{ab} j_{0c}  - {F^\gamma}_{ab} k_\gamma (\sigma) - H_{abc} j_1^c \right) \delta(\sigma-\sigma^\prime) \nonumber \\
\{ j_{0a}(\sigma) , j_1^b(\sigma^\prime) \} &= \delta_a^b \delta^\prime (\sigma - \sigma^\prime) - {f^b}_{ca} j_1^c (\sigma) \delta(\sigma-\sigma^\prime) \nonumber \\
\{j_{0a}(\sigma),k^\beta(\sigma^\prime) \}&= - {F^\beta}_{ca} j_1^c \delta(\sigma-\sigma^\prime) \label{eq:CurrentAlgebraDecomposed}\\
\{ k^\alpha (\sigma), k^\beta (\sigma^\prime) \} &= \kappa^{\alpha \beta} \delta^\prime (\sigma - \sigma^\prime) \nonumber \\
\{ j_1^a (\sigma) , j_1^b(\sigma^\prime) \} &= 0 = \{ j_1^a (\sigma) , k^\beta(\sigma^\prime) \}.  \nonumber
\end{align}
The resulting equations of motion $\partial_\tau \mathcal{O} = \{ \mathcal{O} , H \}$ with the Hamiltonian \eqref{eq:HamiltonianMetricFrame} in the generalised metric frame,
\begin{equation}
	H = \frac{1}{2} \int \mathrm{d} \sigma \ \left( \kappa^{ab} j_{0a}(\sigma) j_{0b}(\sigma) + \kappa_{ab} j_1^a(\sigma) j_1^b(\sigma) + \kappa_{\alpha \beta} k^\alpha(\sigma) k^\beta(\sigma) \right) , \label{eq:HamiltonianFluxFrame}
\end{equation}
are
\begin{align}
0&= \partial_+ j_{-a} + \partial_- j_{+a} - 2 {F^\beta}_{ca} k_\beta j_-^c - H_{abc} j_+^b j_- ^c ,  \nonumber \\
0&= \partial_+ j_-^a - \partial_- j_+^a + {f^a}_{bc} j_+^b j_-^c ,  \label{eq:EquationsOfMotion} \\
0&= 2 \partial_- k^\alpha - {F^\alpha}_{bc} j_+^b j_-^c , \nonumber
\end{align}
in conventions with $V_\pm = V_0 \pm V_1$. The generalisation of these equations of motion to a generic $F_{\mathcal{ABC}}$, corresponding to choices of frame than \eqref{eq:GeneralisedVielbeinGeometric}, is straightforward as for O$(d,d)$, e.g. in \cite{Osten:2019ayq}.

\section{The $A$-deformation of principal chiral model} \label{chap:Adeformation}
There is a plethora of integrable deformations of the principal chiral model: $\lambda$-, $\eta$-, homogeneous Yang-Baxter and WZ-term-deformation and combinations of those. Here, a new deformation of a $d$-dimensional principal chiral model by the heterotic gauge field $A$ is proposed. 

\subsection{Integrability of the undeformed model}
As starting point we take the principal chiral model on a group $G$ with Lie algebra $\mathfrak{g}$, embedded into the bosonic heterotic $\sigma$-model. In the language of the previous section, this will be characterised by the components of the Maurer-Cartan form on $G$ as vielbein $e$ and $B = A = 0$ in the generalised vielbein \eqref{eq:GeneralisedVielbeinGeometric}. This corresponds to the generalised fluxes \eqref{eq:GeneralisedFluxes}
\begin{equation}
	{f^c}_{ab}, \qquad H_{abc} = 0, \qquad {F^\gamma}_{ab} = 0,
\end{equation}
where ${f^c}_{ab}$ are structure constants to the Lie algebra $\mathfrak{g}$. This corresponds to the following equations of motion
\begin{equation}
\partial_+ j_{-a} + \partial_- j_{+a}= 0 , \qquad \partial_+ j_-^a - \partial_- j_+^a + {f^a}_{bc} j_+^b j_-^c= 0, \qquad \partial_- k ^\alpha = 0 . \label{eq:EquationsOfMotionUndeformed}
\end{equation}
This model is classically integrable, as the currents $k$ and the $d$-dimensional principal chiral model with the currents $j$ decouple. Classical integrability is shown by providing a \textit{Lax representation} and a classical $\mathcal{R}$-matrix, proving Hamiltonian integrability, i.e. involution of the charges produced from the Lax representation.

The Lax representation of this model is
\begin{equation}
L_+(\lambda) =  \frac{1}{1 - \lambda} \big( j_+^a t_a + 2 k^\alpha t_\alpha \big), \qquad L_-(\lambda) = \frac{1}{1+\lambda} j_-^a t_a \qquad \in \tilde{\mathfrak{g}} = \mathfrak{g} \oplus \mathfrak{u}(1)^n , \label{eq:LPairUndeformed}
\end{equation}
where $\lambda$ is the spectral parameter. $(t_a , t_\alpha)$ are generators of $\tilde{\mathfrak{g}} = \mathfrak{g} \oplus \mathfrak{u}(1)^n$ which is defined by
\begin{equation}
	[t_a , t_b] = {f^c}_{ab} t_c, \qquad [t_\alpha , t_\beta ] = 0 = [t_a , t_\alpha ]. \nonumber 
\end{equation} 
A method to show involution of the associated tower of conserved charges is Maillet's formalism \cite{Maillet:1985ec,Maillet:1986}. Central for this is the \textit{Lax matrix}
\begin{equation}
\mathcal{L}(\lambda) = L_1(\lambda) = \frac{1}{2} ( L_+(\lambda) - L_-(\lambda) ) = \frac{\lambda j_0^a + j_1^a }{1 - \lambda^2} t_a + \frac{1}{1-\lambda} k^\alpha t_\alpha
\end{equation}
The aim is to construct a so-called classical $\mathcal{R}$-matrix, i.e. an object $\mathcal{R}(\lambda,\mu) \in \tilde{ \mathfrak{g}} \otimes \tilde{\mathfrak{g}}$ such that
\begin{align}
\{ \mathcal{L}_{\mathbf{1}}(\lambda,\sigma), \mathcal{L}_{\mathbf{2}}(\mu,\sigma^\prime) \} &= \left( [ \mathcal{R}_{\mathbf{12}}(\lambda,\mu),\mathcal{L}_{\mathbf{1}}(\lambda,\sigma)] - [ \mathcal{R}_{\mathbf{21}}(\mu,\lambda),\mathcal{L}_{\mathbf{2}}(\mu,\sigma)] \right) \delta(\sigma - \sigma^\prime) \nonumber \\
&{} \quad - \left( \mathcal{R}_{\mathbf{12}}(\lambda,\mu) + \mathcal{R}_{\mathbf{21}}(\mu,\lambda) \right) \delta^\prime(\sigma - \sigma^\prime) .  \label{eq:MailletBracket}
\end{align}
A bold subscript indicates the place in the tensor product, e.g. $X_{\mathbf{1}} = X \otimes \mathbb{1}$. A sufficient condition for the involution of the conserved charges constructed from the Lax pair is the \textit{classical Yang-Baxter equation}:
\begin{equation}
[ \mathcal{R}_{\mathbf{12}}(\lambda_1 , \lambda_2) , \mathcal{R}_{\mathbf{13}}(\lambda_1 , \lambda_3) ] + [ \mathcal{R}_{\mathbf{12}}(\lambda_1 , \lambda_2) , \mathcal{R}_{\mathbf{23}}(\lambda_2 , \lambda_3) ] + [ \mathcal{R}_{\mathbf{32}}(\lambda_3 , \lambda_2) , \mathcal{R}_{\mathbf{31}}(\lambda_3 , \lambda_1) ] = 0 . \label{eq:CYBE}
\end{equation}
The classical $\mathcal{R}$-matrix for the undeformed model \eqref{eq:EquationsOfMotionUndeformed} is
\begin{equation}
\mathcal{R}(\lambda,\mu) = \frac{t_a \otimes t^a}{\lambda - \mu} \phi^{-1}_{PCM} (\mu) + \frac{t_\alpha \otimes t^\alpha}{\lambda-\mu} \phi^{-1}_{chiral}(\mu). \label{eq:RMatrixUndeformed}
\end{equation}
with the so-called twist functions
\begin{equation}
\phi_{PCM} (\lambda) = \frac{1 - \lambda^2}{\lambda^2}, \qquad \phi_{chiral} (\lambda) = 1 - \lambda.
\end{equation}
This is close to the \textit{twist function form} $\mathcal{R}(\lambda,\mu) = \mathcal{R}^{(0)}(\lambda,\mu) \phi^{-1}(\lambda)$ with a solution of the classical Yang-Baxter equation $\mathcal{R}^{(0)}$, like $\mathcal{R}(\lambda,\mu) = \frac{t_a \otimes t^a}{\lambda - \mu}$ and a twist function $\phi(\mu)$. This form is a sufficient condition in order for $\mathcal{R}$ to be a solution of the classical Yang-Baxter equation \eqref{eq:CYBE} and a typical assumption in the classification of integrable $\sigma$-models \cite{Lacroix:2023gig}. From this point of view it is noteworthy that this simple model has no $\mathcal{R}$-matrix of this form. Nevertheless, it is easy to see that \eqref{eq:RMatrixUndeformed} solves the classical Yang-Baxter equation, as $\mathfrak{t}_\alpha \otimes t^\alpha$ is a central element of $\tilde{\mathfrak{g}} \otimes \tilde{\mathfrak{g}}$, and all its contributions drop out of the classical Yang-Baxter equation \eqref{eq:CYBE}.

\subsection{Introducing the deformation}

The deformation is characterised by a constant (in flat indices) heterotic gauge field 
\begin{equation}
	A_a^\alpha = \epsilon \mathcal{A}_a^\alpha, \quad \text{with} \quad  \mathcal{A}_a^\alpha \mathcal{A}_b^\beta \kappa_{\alpha \beta} = \kappa_{ab}, \label{eq:DefA}
\end{equation}
so $\mathcal{A} \in \ $O$(d) \subset \ $Mat$(n,d)$. 
For this one needs that $n > d$ which would be the case for the bosonic part of the heterotic string where $n=16$.\footnote{This is due the fact that U$(1)^n$ is the Cartan subgroup of $E_8 \times E_8$ or SO$(32)$.} We use $\kappa_{ab}$ (as the Killing metric on $\mathfrak{g})$ and $\kappa_{\alpha \beta}$ to raise and lower indices on $\mathcal{A}$, e.g. to define the object $\mathcal{A}^a_\alpha$. Care needs to be taken that when contracting $\mathcal{A}$'s in the $n$-dimensional indices one will produce $A_a^\alpha \mathcal{A}_{b\alpha} = \kappa_{ab} $ by definition \eqref{eq:DefA}, but contractions of the $d$-dimensional flat indices will result in a projector on the part of the $n$-dimensional currents, that couple to the $j_+$ (see below in the equations of motions) 
\begin{equation}
	\mathcal{P}^{\alpha \beta} = \mathcal{A}^\alpha_a \mathcal{A}^{a \beta}. 
\end{equation}
In particular, one has that $ {\mathcal{P}^\alpha}_\gamma \mathcal{P}_{\gamma \beta} = \mathcal{P}_{\alpha \beta}$ and ${\mathcal{P}^\alpha}_\beta \mathcal{A}^\beta_a = \mathcal{A}^\alpha_b$. So contractions like $\mathcal{A}_{a\alpha} k^\alpha$ will project out the decoupled components of $k$ as well. One could, of course, decompose the $\mathfrak{u}(1)^n$-currents into coupled and decoupled currents immediately, by choosing the generators of $\mathfrak{u}(1)^n$ accordingly. Also, in this sense, there is only \textit{one} inequivalent $A$-deformation. All $\mathcal{A} \in \ $O$(d)$ are related by a basis change of $\mathfrak{u}(1)^n$. Keeping in mind possible generalisations, the $n$-dimensional covariance by introducing $\mathcal{A}$ is kept in the following.

Resultingly, any choice of gauge field satisfying \eqref{eq:DefA}, should result in the same deformation of the fluxes
\begin{align}
	H_{abc} &= \left( \tilde{h} - \frac{3}{2} \epsilon^2 \right) f_{abc} = h f_{abc}, \nonumber \\
	{f^c}_{ab} &= {f^c}_{ab}, \label{eq:FluxesDeformation} \\
	{F^\gamma}_{ab} &= - \epsilon \mathcal{A}_c^\gamma {f^c}_{ab} , \nonumber
\end{align}
where an WZ-term with some free coefficient $\tilde{h}$ was allowed as well, i.e. a $B$-field such that $H \sim f$. Effectively, this results in a \textit{two-parameter deformation} where the WZ-part is, of course, well understood but it interesting to note that one can combine it with the novel $A$-deformation. The equations of motion are:
\begin{align}
	0 &= \partial_+ j_-^a + \partial_- j_+^a + 2 \epsilon \mathcal{A}^b_\beta {f^a}_{bc} k_+^\beta j_-^c - h {f^a}_{bc} j_+^b j_-^c , \nonumber \\
	0 &= \partial_+ j_-^a - \partial_- j_+^a + {f^a}_{bc} j_+^b j_-^c  ,
	 \label{eq:EquationsOfMotionDeformation} \\
	0 &=  2 \partial_- k^\alpha + \epsilon \mathcal{A}^\alpha_a {f^a}_{bc} j_+^b j_-^c . \nonumber
\end{align}
It is convenient to decompose the last equation into an $d$-dimensional part of $k$ that couples to the $j$, and the $n-d$-dimensional part that is still decoupled:
\begin{align}
0 &= \mathcal{A}^a_\alpha \left( 2 \partial_- k^\alpha + \epsilon {f^a}_{bc}j_+^b j_-^c \right),  \qquad  0 = (1 - P)_{\alpha \beta} \partial_- k^\beta.
\end{align}
These decoupled components and the associated generators will be denoted with a bar in the following:
\begin{align*}
{(1-P)^\alpha }_\beta k^\beta = \bar{k}^\alpha, \qquad {(1-P)_\alpha }^\beta t_\beta = \bar{t}_\alpha.
\end{align*} 

\subsection{The Lax pair}
In order to show the existence of a Lax pair $L_\pm \in \tilde{\mathfrak{g}}$, we make the ansatz is
\begin{align}
	L_-(\lambda) = a(\lambda) j_-^a t_a, \qquad L_+ = \big( b(\lambda) j_+^a + c(\lambda) \epsilon \mathcal{A}^a_\alpha k^\alpha \big) t_a + \frac{2}{1-\lambda} \bar{k}^\alpha \bar{t}_\alpha .
\end{align}
where the coefficient in front of the decoupled currents $\bar{k}$ is taken from the undeformed case \eqref{eq:LPairUndeformed} and $\lambda$ is the spectral parameter. The coefficients $a(\lambda),b(\lambda),c(\lambda)$ are determined such that
\begin{equation}
	0= \partial_+ L_- - \partial_- L_+ + [L_+, L_-] \in \tilde{\mathfrak{g}}
\end{equation}
is valid by virtue of the equations of motion \eqref{eq:EquationsOfMotionDeformation}. This results in the under-determined system of equations:
\begin{align*}
0= \frac{a+b}{2} - ab - \frac{1}{2} ( \epsilon^2 c + h (a-b) ), \qquad 0= - (a-b) + ac.
\end{align*}
A convenient parametrisation of the solution to this is
\begin{equation}
	a = \frac{1}{1+\lambda}, \quad b= \frac{1- [ \epsilon^2 (1+\lambda) + h ] }{f_{\epsilon,h}(\lambda)}, \quad c = - \frac{2\lambda}{f_{\epsilon,h}(\lambda)},
\end{equation}
with $f_{\epsilon,h}(\lambda) =  1 - \lambda - (1+\lambda) [ \epsilon^2 (1+\lambda) + h ]$. In that parametrisation one can recognise a typical parametrisation of the Lax pair of the principal chiral model in the undeformed limit. I.e. for the $\mathfrak{g}$-valued part of $L(\lambda)$
\begin{equation}
	L^a_\pm (\lambda) \longrightarrow \frac{j_\pm^a}{1 \mp \lambda}, \quad \text{for} \quad \epsilon,h \rightarrow 0.
\end{equation}
But the full Lax pair $L_\pm \in \tilde{\mathfrak{g}} $ in the limit $\epsilon \rightarrow 0$ does \textit{not} give the full undeformed Lax pair \eqref{eq:LPairUndeformed} -- the part of the $k^\alpha$ that is coupled in the deformed models disappears from the Lax pair and the equations of motion reconstructed from it.

\subsection{Hamiltonian integrability}
One should check whether the conserved charges from the Lax pair are also in Poisson involution for arbitrary choices of $\epsilon$ and $h$. The relevant quantity for this is the Lax matrix
\begin{align}
	\mathcal{L} &\equiv L_1 = \frac{1}{(1+\lambda) f_{\epsilon,h}(\lambda)} \left[ \lambda j_0^a + \left( f_{\epsilon,h}(\lambda) + \lambda) \right) j_1^a - \epsilon \ \lambda (1+ \lambda) \mathcal{A}_\alpha^a k_+^\alpha \right] t_a + \frac{1}{1-\lambda} \bar{k}^\alpha \bar{t}_\alpha . \label{eq:LaxMatrix}
\end{align}
A corresponding classical $\mathcal{R}$-matrix is
\begin{equation}
\mathcal{R}_{\mathbf{12}}(\lambda,\mu) = \frac{t_a \otimes t^a}{\lambda - \mu} \phi^{-1}_{\epsilon,h}(\mu) + \frac{\bar{t}^\alpha \otimes \bar{t}_\alpha}{\lambda - \mu} \frac{1}{1 - \mu}, \qquad \text{with} \quad \phi_{\epsilon,h}(\mu) = \frac{(1+\mu)f_{\epsilon,h}(\mu)}{\mu^2} . \label{eq:RMatrix}
\end{equation}
Again, $\mathcal{R}$ is almost of the form with twist function and solves the classical Yang-Baxter equation \eqref{eq:CYBE} by the same argument as in the undeformed case. This $\mathcal{R}$-matrix could be guessed relatively easily from inspiration from the form of the $\mathcal{R}$-matrix of the principal chiral model. It is easy to see that the $\mathfrak{g}\otimes \mathfrak{g}$-valued part is reproduced for $\epsilon,h \rightarrow 0$, as then $f_{\epsilon,h}(\lambda) \rightarrow (1 - \lambda)$. But, as also for the Lax pair, the full $\mathcal{R}$-matrix \eqref{eq:RMatrix} does not reproduce the full undeformed $\mathcal{R}$-matrix \eqref{eq:RMatrixUndeformed} in the non-deformation limit.

One checks that \eqref{eq:RMatrix} and \eqref{eq:LaxMatrix} fit into Maillet's formalism, by showing that for the $\mathfrak{g} \otimes \mathfrak{g}$-valued part both the left-hand and right-hand side of Maillet bracket \eqref{eq:MailletBracket} correspond to
\begin{align}
&{} \qquad \text{\scriptsize$\frac{ t^a \otimes t^b}{(1 + \lambda) f_{\epsilon,h}(\lambda) (1 + \mu) f_{\epsilon,h}(\mu) }$} \Big( {f^c}_{ab} \ \delta(\sigma - \sigma^\prime) \ \Big( ( - \lambda \mu ) j_{0c}(\sigma) + ( \lambda \mu ) \ \epsilon \mathcal{A}^\gamma_c k_\gamma(\sigma)  \nonumber \\
&{} \qquad \qquad - \big[ (\lambda + \mu) - \epsilon^2 ( \lambda + \mu + 2 \lambda \mu ) - h ( \lambda + \mu + \lambda \mu ) \big]  j_{1c}(\sigma) \Big)  \label{eq:RMatrixCheck} \\
&{} \qquad \qquad + \kappa_{\alpha\beta} \delta^\prime(\sigma - \sigma^\prime) \  \Big( (\lambda + \mu) - \epsilon^2 ( \lambda + \mu  + 3 \mu \lambda - \lambda^2 \mu^2 ) - h ( \lambda + \mu  + 2\mu \lambda ) \Big) \Big). \nonumber
\end{align}
The part valued in $\mathfrak{u}(1)^n \otimes \mathfrak{u}(1)^n$ remains unchanged from the discussion in the undeformed case.

\section{Discussion}
In this article a new integrable deformation of the principal chiral model inspired by the heterotic string was proposed. It builds on the recent Hamiltonian formulation of the bosonic part of the heterotic string $\sigma$-model in terms of $O(d,d+n)$-generalised geometry in \cite{Hatsuda:2022zpi}. Building on that the analogue to the so-called $\mathcal{E}$-model in O$(d,d)$-generalised geometry was proposed in section \ref{chap:Review}. The deformation was introduced in section \ref{chap:Adeformation} and is characterised by a certain choice of heterotic gauge field $A^\alpha = \epsilon \mathcal{A}^\alpha$. The concrete choice of $\mathcal{A} \in \ $O$(d)$, can be absorbed into a change of basis for the $n$-dimensional current $k$. Hence, this deformation of $A$ really only results in a one-parameter deformation given by the fluxes \eqref{eq:FluxesDeformation}. It is also shown that this deformation can be extended to a two-parameter deformation together with a WZ-term like deformation, while preserving integrability. Classical integrability was shown by the explicit construction of a Lax pair \eqref{eq:LaxMatrix} and a classical $\mathcal{R}$-matrix \eqref{eq:RMatrix} that solves the classical Yang-Baxter equation.

To the knowledge of the author this is uncharted territory. From here several direction should be taken to further explore this space of novel integrable $\sigma$-models:
\begin{itemize}
\item Generalisation to the full (supersymmetric) heterotic string
\item Different starting points for the deformation: 
\begin{itemize}
\item[-] symmetric space coset $\sigma$-models, Yang-Baxter or $\lambda$-deformation, instead of the principal chiral model for a Lie group $G$

\item[-] non-abelian chiral currents $k^\alpha$, i.e. some non-trivial gauge algebra $\mathfrak{h}$ instead of $\mathfrak{u}(1)^n$
\end{itemize}

\item Dual version of the $A$-deformation.\footnote{In generalisation of the fact, that one can have integrable deformations based on the Kalb-Ramond field $B$ (WZ-term) or their duals $\beta$ ($\eta$- and homogeneous Yang-Baxter deformations), there could be also a dual version of $A$-deformation.} The natural starting point would be the dual vielbein to \eqref{eq:GeneralisedVielbeinGeometric} with dual fields $\alpha$ and $\beta$ to $A$ and $B$:
\begin{equation}
	{E_\mathcal{A}}^\mathcal{M} = \left( \begin{array}{ccc} \delta_a^b & 0 & 0 \\ \alpha_\alpha^b & \delta_\alpha^\beta & 0 \\
	 \beta^{ab} - \frac{1}{2} \alpha^a_\alpha \alpha^{b \alpha} & - \alpha^{a\beta} & \delta^a_b
	\end{array} \right) \left( \begin{array}{ccc} {e_b}^m & 0 & 0 \\ 0 & {\delta_\beta}^\mu & 0 \\ 0 & 0 & {e^b}_m	
	\end{array} \right). \label{eq:GeneralisedVielbeinNonGeometric}
\end{equation}
\item the deformed models as reduction of known integrable $\sigma$-models.

From the point of view of the systematics of integrable models, the question is: how new is this model really? The present Lax representation is valued in $\mathfrak{g} \oplus \mathfrak{u}(1)^n$. A natural candidate for a more conventional integrable $\sigma$-model, that would reproduce the same equations of motion \eqref{eq:EquationsOfMotion}, would be something like a principal chiral model or a WZW-model with target space $G \times T^n$, together with some constraint that makes $k_-$ (which did not appear in the construction here) vanish. This is clearly possible for the undeformed model, but does not seem possible straightforward for the deformed model. This deserves further attention.

A related question would be whether these models can be obtained from 4d Chern-Simons theory \cite{Costello:2017dso,Costello:2018gyb,Bittleston:2020hfv,Lacroix:2020flf,Levine:2023wvt,Cole:2023umd}.  The not quite 'twist function'-form of the $\mathcal{R}$-matrices \eqref{eq:RMatrixUndeformed} and \eqref{eq:RMatrix} might suggest that this is not the case.

\end{itemize}
Beside these questions, two issues should be addressed and will be briefly discussed below -- the questions about the embedding of these models into string theory and their incorrect non-deformation limit. Both questions certainly require more study, in particular if the present approach can be extended to the above mentioned more non-trivial examples. 

\subsection{Embedding into heterotic supergravity} 
The usual question for the integrable models and their deformations in context of string theory is, whether they are also supergravity backgrounds. A detailed discussion is left for future research when a similar analysis to the one in this article has been done with the inclusion of fermions. 

This question consists of two parts:
1) When is the undeformed model a solution of heterotic supergravity. 2) Is the $A$-deformation a solution to heterotic supergravity, given the undeformed model is?

\paragraph{The principal chiral model.} In comparison to the type II $\sigma$-models, the heterotic principal chiral model \textit{cannot} be completed a string $\sigma$-model by simply choosing a supergroup with vanishing Killing form \cite{Bershadsky:1999hk}.

\paragraph{The WZW-type undeformed model.} A valid starting point for the deformation in heterotic supergravity would be the WZW-type model, i.e. target space geometry corresponding to the fluxes \eqref{eq:FluxesDeformation} with $h=1$, $\epsilon = 0$. This is clearly a solution of heterotic supergravity, as it only contains NS-NS fields and is a conformal model. WZW-models have been applied in several circumstances in heterotic string theory, for example for the heterotic string on AdS$_3$ \cite{Hohenegger:2008du} or compactifications on group manifolds with torsion \cite{Chatzistavrakidis:2009mh}. Some of these discussions actually require (chiral) gauged WZW-models \cite{Gannon:1992np,Chung:1992mj,Sfetsos:1993bh}. For this, the present approach has to be extended from group to coset manifolds as discussed above. 

\paragraph{The deformed model.} One can check that an integrable deformation of such a WZW-model will correspond to a solution of heterotic supergravity, given that the two deformation parameters $h$ and $\epsilon$ in \eqref{eq:FluxesDeformation} are related in a certain way. 

For example, one of the equations of motion of heterotic supergravity \cite{Sen:1985qt} -- in the frame formalism and for constant dilaton -- is:
\begin{equation}
0 = R^{(+)}_{ab} - 2\alpha^\prime \left( \mathrm{tr} (F_{ac} {F_b}^c) - R^{(+)}_{ac}  {{R^{(+)}}_b}^c \right). \label{eq:HeteroticSugra}
\end{equation}
Here, $R^{(+)}$ is, as usual, the Ricci tensor of the torsionful connection $\Gamma + \frac{1}{2} H$, where $\Gamma$ is the Christoffel connection. The choice of fluxes \eqref{eq:FluxesDeformation} means that
\begin{align*}
\mathrm{tr} (F_{ac} {F_b}^c) &= \epsilon^2 \kappa_{ab}, \qquad R_{ab}^{(+)} = - \chi^2 \kappa_{ab}
\end{align*}
with the Killing form $\kappa$ on $\mathfrak{g}$ and $\chi^2 = \frac{1}{4} (h-1)^2 ( 2 + (h-1)^2)$ for simplicity of notation.

So far, we ignored dimensions. Only for the following analysis, these are briefly reintroduced. In particular, $ [\chi^2] = [\epsilon] = \left[ \alpha^\prime \right]^{-1}$. The heterotic supergravity equation \eqref{eq:HeteroticSugra} at $\mathcal{O}(\alpha^\prime)$ is satisfied, given that deformation parameters are related: 
\begin{equation}
\chi^2 = \frac{1}{4\alpha^\prime} - \sqrt{\frac{1}{16 \alpha^{\prime 2}} + \epsilon^2} \approx 2 \alpha^\prime \epsilon^2 + (\alpha^{\prime})^{-1} \mathcal{O}\left((\alpha^\prime \epsilon)^4 \right) .
\end{equation}
The other heterotic supergravity equations hold in a similar way. Also, by virtue of the Jacobi identity of $\mathfrak{g}$, the anomaly cancellation condition
\begin{equation}
\mathrm{d} H = 0 = \mathrm{tr} ( F \wedge F )
\end{equation}
is automatically fulfilled for fluxes of the deformed model \eqref{eq:FluxesDeformation}.

\subsection{Non-deformation limit.}
As discussed briefly in the main text, the objects of the Lax construction, i.e. the Lax pair and the classical $\mathcal{R}$-matrix do not have the correct non-deformation limit. The $\mathfrak{g}$-valued part of the Lax connection \eqref{eq:LaxMatrix} correctly reproduces the one of the principal chiral model, but the full $\mathfrak{g}\oplus \mathfrak{u}(n)^n)$-valued Lax connection does not reproduce the full undeformed one \eqref{eq:LPairUndeformed}. A similar problem also arises for the poles and zeros of the twist function.
Whereas $\phi_{\epsilon,h} \rightarrow \phi_{PCM}$ and both have double poles at $0$ and infinity, $\phi_{\epsilon,h} = (1+ \lambda) f_{\epsilon,h}(\lambda)\lambda^{-2}$ has three zeros:
\begin{equation}
\lambda_1 = -1 , \qquad \lambda_{2,3} = - \frac{1}{2 \epsilon^2} \left( 2\epsilon^2 + (1+h) \pm \sqrt{8 \epsilon^2 + (h+1)^2} \right)
\end{equation}
For an analysis of the $h,\epsilon \rightarrow 0$-limit, a concrete $\epsilon$-dependence of $h = \tilde{h} - \frac{3}{2}\epsilon^2$ resp. also of $\tilde{h}$ needs to be specified. As an example, let us assume that $h = h_0$ has no $\epsilon$-dependence (so $\tilde{h} = h_0 + \frac{3}{2} \epsilon^2$). Then one of the $\lambda_{2,3}$ has the expected $\epsilon \rightarrow 0$-limit:
\begin{align}
\lambda_2 = - \frac{1+h_0}{\epsilon^2} - \frac{3}{1+h_0} + \mathcal{O}(\epsilon^2) \overset{\epsilon,h_0 \rightarrow 0}{\longrightarrow} - \infty, \qquad \lambda_3 = \frac{1}{1+h_0} + \mathcal{O}(\epsilon^2) \overset{\epsilon,h_0 \rightarrow 0}{\longrightarrow} = 1
\end{align}
Latter gives the expected zero in comparison to the twist function of the principal chiral model, but the role and origin of the additional zero at infinity is not clear at this stage. Let us summarise these issues:
\begin{align}
	\text{deformed Lax pair \eqref{eq:LaxMatrix}} &\nrightarrow \text{undeformed Lax pair \eqref{eq:LPairUndeformed}} \nonumber \\
	\text{deformed $\mathcal{R}$-matrix \eqref{eq:RMatrix}} &\nrightarrow \text{undeformed $\mathcal{R}$-matrix \eqref{eq:RMatrixUndeformed}} \qquad \text{for} \ \epsilon,h \rightarrow 0 \\
	\text{3 zeros of $\phi_{\epsilon,h}$} &\nrightarrow \text{2 zeros of $\phi_{PCM}$}. \nonumber
\end{align}
This result might seem puzzling, as for the known integrable deformations the non-deformation limit gives the expected undeformed Lax pairs, see e.g. \cite{Hoare:2021dix}. An explanation for this might be the following: It is well-known in perturbation theory that deformations (or perturbations) of partial differential equations do not necessarily correspond to deformations of the solution space. One might interpret this result in this context, as the Lax pair corresponds to a solution of the system. Nevertheless, this question deserves further attention.

\section*{Acknowledgements}
The author thanks Falk Hassler, Eric Lescano and the anonymous referee for PRD for helpful comments on the draft.

This research is part of the project No. 2022/45/P/ST2/03995 co-funded by the National Science Centre and the European Union’s Horizon 2020 research and innovation programme under the Marie Sk\l odowska-Curie grant agreement no. 945339.

\vspace*{10pt}
\includegraphics[width = 0.09 \textwidth]{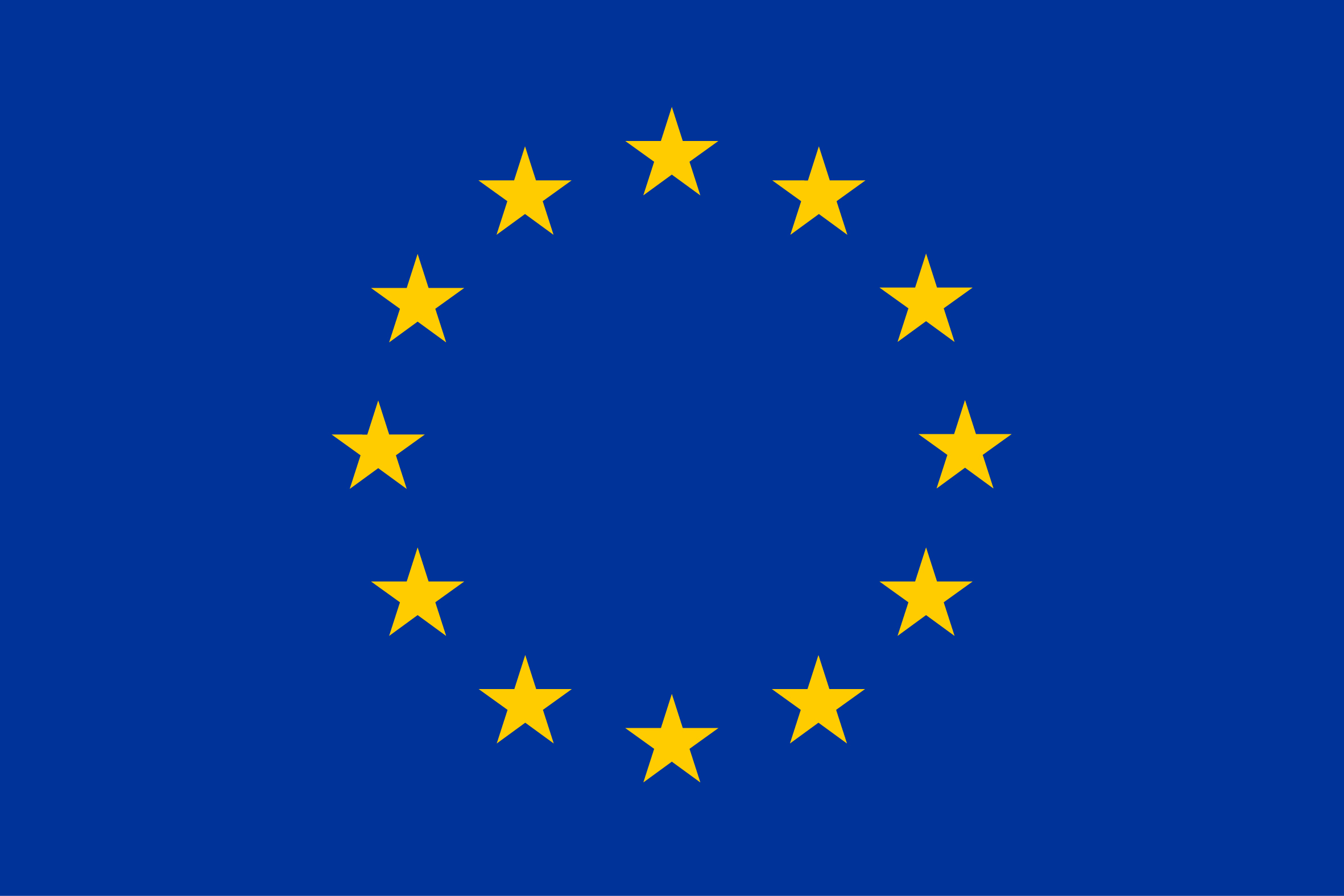} $\quad$
\includegraphics[width = 0.7 \textwidth]{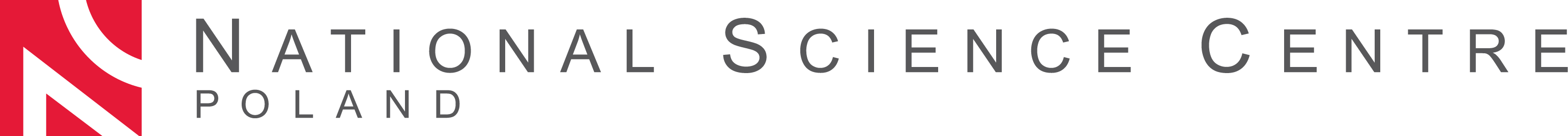}

\bibliographystyle{jhep}
\bibliography{References}
\end{document}